\documentstyle[epsfig,psfig]{mn2e}

\title[The Nature of Faint Submm Galaxies]{The SCUBA Half-Degree Extragalactic Survey (SHADES) - VIII. The Nature of Faint Submm Galaxies in SHADES, SWIRE and SXDF Surveys}
\author[D.L. Clements et al.]
       {D.L. Clements$^1$,  M.Vaccari$^1$, T. Babbedge$^1$, S. Oliver$^2$, M. Rowan-Robinson$^1$, \newauthor P. Davoodi$^2$, R. Ivison$^{3,4}$, D. Farrah$^5$, J. Dunlop$^3$, Dave Shupe$^{22}$, I. Waddington$^2$, \newauthor C. Simpson$^{7}$, H. Furusawa$^{20}$, S. Serjeant$^9$, A. Afonso-Luis$^{23}$, D.M. Alexander$^{10, 8}$, \newauthor  I. Aretxaga$^{11}$, A. Blain$^{12}$,  C. Borys$^{12}$,  S. Chapman$^{12}$, K. Coppin$^{8}$, L. Dunne$^{3,13}$, S. Dye$^{14}$,   \newauthor S.A. Eales$^{14}$, T. Evans$^6$, F. Fang$^{22}$, D. Frayer$^6$, M. Fox$^1$, W.K. Gear$^{14}$, T.R. Greve$^{12}$,  \newauthor M. Halpern$^{15}$, D.H. Hughes$^{11}$, T. Jenness$^{16}$,  C.J. Lonsdale$^{6,21}$, A.M.J. Mortier$^3$,  \newauthor M.J. Page$^{17}$, A. Pope$^{15}$, R.S. Priddey$^{18}$, S. Rawlings$^{19}$, R.S. Savage$^2$,  \newauthor D. Scott$^{15}$, S.E. Scott$^{3}$, K. Sekiguchi$^{20,26}$,  I. Smail$^8$, H.E. Smith$^{21}$,  J.A. Stevens$^{18}$, \newauthor J. Surace$^{22}$, T. Takagi$^{25}$, E. van Kampen$^{24}$\\
         $^1$Imperial College, London, Blackett Lab, Prince Consort Road, London SW7 2BW, UK\\
         $^2$Astronomy Centre, University of Sussex, Falmer, Brighton BN1 9QH\\
         $^3$ SUPA\footnote{Scottish Universities Physics Alliance} Institute for Astronomy, University of Edinburgh, Royal Observatory, Edinburgh EH9 3HJ\\
	$^4$UK ATC, Royal Observatory, Blackford Hill, Edinburgh EH9 3HJ\\
         $^5$Department of Astronomy, Cornell University, Space Sciences Building, Ithaca, NY 14853, USA\\
         $^6$IPAC, MS 100-22, Caltech, JPL, Passadena, CA 91125, USA\\
         $^7$Astrophysics Research Institute, Liverpool John Moores University, Twelve Quays House, Egerton Wharf, Birkenhead CH41 1LD\\
         $^8$Institute for Computational Cosmology, Department of Physics, Durham University, Durham DH1 3LE\\
         $^9$CEPSR, Open University, Milton Keynes\\
         $^{10}$Institute of Astronomy, University of Cambridge, Madingley Road, Cambridge, CB3 0HA\\
         $^{11}$Instituto Nacional de Astrofisica, Optica y Electronica, Apartado Postal 51 y 216, 72000, Puebla, Pue, Mexico\\
         $^{12}$Caltech, 1200 E California Blvd., Passadena, CA 91125-0001, USA\\
         $^{13}$The School of Physics and Astronomy, University of Nottingham, University Park, Nottingham NG7 2RD\\
         $^{14}$School of Physics and Astronomy, Cardiff University, 5 The Parade, Cardiff CF24 3YB\\
         $^{15}$Department of Physics and Astronomy, University of British Columbia, 6224 Agricultural Road, Vancouver, BC V6T 1Z1, Canada\\
         $^{16}$Joint Astronomy Centre, 660 N. A'ohoku Place, University Park, Hilo, HI 96720 USA\\
         $^{17}$Mullard Space Science Laboratory (MSSL), University College London, Holmbury St Mary, Dorking, Surrey RH5 6NT\\
         $^{18}$Centre for Astrophysics Research, Science and Technology Research Institute, University of Hertfordshire, College Lane, Hatfield,\\ Hertfordshire AL10 9AB\\
         $^{19}$Department of Astrophysics, Denys Wilkinson Building, Keble Road, Oxford OX1 3RH\\
         $^{20}$Subaru Telescope, National Optical Astronomy Observatory of Japan, 650 N. A'ohoku Place, Hilo, HI 96720, USA\\
         $^{21}$Centre for Astrophysics and Space Sciences, University of California San Diego, La Jolla, CA 92063-0424, USA\\
         $^{22}$Spitzer Science Centre, MS 220-6, Caltech, Jet Propulsion Lab, Pasadena, CA 91125, USA\\
         $^{23}$Instituto de Astrofisica de Canarias, La Laguna, Spain\\
         $^{24}$Institute for Astro- and Particle Physics, University of Innsbruck, Technikerstr. 25, A-6020 Innsbruck, Austria\\
	$^{25}$Institute of Space and Astronautical Science, Japan Aerospace
Exploration Agency, Sagamihara, Kanagawa 229-8510, Japan\\
	$^{26}$National Astronomical Observatory of Japan, Mitaka, Tokyo 181-8588, Japan\\
\\
\\
\\
\\
\\
\\
\\
\\
\\
\\
\\
\\
\\
\\
\\
}
\date{}
\pagerange{\pageref{firstpage}--\pageref{lastpage}}
\pubyear{}
\begin{document}
\label{firstpage}
\maketitle
\begin{abstract}
We present the optical-to-submm spectral energy distributions for 33 radio \& mid-IR identified
submillimetre galaxies  discovered via the SHADES 850${\rm \mu m}$ SCUBA imaging
in the Subaru-XMM Deep Field (SXDF). Optical data for the sources comes from the Subaru-XMM Deep Field (SXDF) and mid- and far-IR fluxes from SWIRE.
We obtain photometric redshift estimates for our sources using optical and IRAC 3.6 and 4.5 $\mu$m fluxes. We then fit spectral energy distribution (SED) templates to the longer wavelength data to determine the nature of the far-IR emission that dominates the bolometric luminosity of these sources. The infrared template fits are also used to resolve ambiguous identifications and cases of redshift
aliasing.  The redshift distribution obtained broadly matches previous results for submm sources and on the SHADES
 SXDF field. Our template fitting finds that AGN, while present in about 10\% of our sources, do not contribute significantly to their bolometric luminosity. Dust heating by starbursts, with either Arp220 or M82 type SEDs, appears to be responsible for the luminosity in most sources (23/33 are fitted by Arp220 templates, 2/33 by the warmer M82 templates). 8/33 sources, in contrast, are fit by a cooler cirrus dust template, suggesting that cold dust has a role in some of these highly luminous objects. Three of our sources appear to have multiple identifications or components at the same redshift, but we find no statistical evidence that close associations are common among our SHADES sources. Examination of 
rest-frame $K$-band luminosity suggests that 'downsizing' is underway in the submm galaxy population, with lower redshift systems lying in lower mass host galaxies. Of our 33 identifications six are found to be of lower reliability but their exclusion would not significantly alter our conclusions.
\end{abstract}


\section{Introduction}

It is now clear that dust enshrouded galaxies play a major role in galaxy evolution. The discovery of the cosmic infrared background (CIB) (Puget et al., 1996; Fixsen et al., 1998) demonstrated that roughly 50\% of the energy generation in the history of the universe took place in dusty, obscured regions, while deep submm surveys (Smail et al., 1997; Hughes et al., 1998; Eales et al., 2000; Scott et al., 2002) have identified brighter members of the population responsible for some of this background. These submillimetre galaxies (SMGs) are thought to be the high redshift equivalents of local ultraluminous infrared galaxies (ULIRGs - Sanders \& Mirabel, (1996); Lonsdale et al. (2006), and references therein), lying at redshifts $z \sim$2.5 (Chapman et al., 2005), hosting starbursts forming stars at 10$^2$ --- 10$^3$ M$_{\odot}$yr$^{-1}$. These objects are likely to be the progenitors of massive elliptical galaxies seen in the local universe today (Blain et al., 2002, \& references therein; Farrah et al., 2006; Takagi et al. 2004).
Despite the rapid development of submm astronomy over the last decade numerous problems remain in our understanding of the sources responsible for the CIB. Many of these problems could be addressed if we had a better idea of the broad band spectral energy distribution (SED) of SMGs, from optical through the mid-IR and into the far-IR.

The role of AGN in the SMG population is currently unclear. It has generally been assumed that the vast majority of the dust heating in these systems, which produces their phenomenal far-IR/submm luminosity, is due to a massive starburst. There is a growing body of evidence, though, that many SMGs might also host a supermassive black hole (SMBH) and an active nucleus. Analysis of the X-ray properties of radio selected, spectroscopically identified SMGs lying in the 2Ms Chandra Deep Field North (CDF-N) by Alexander et al. (2005) show that 75\% of these sources have AGN powered X-ray emission. Many of these sources are heavily obscured (N$_H > 10^{23} {\rm cm^{-2}}$).
Meanwhile, near-IR spectroscopy has shown that at least 40\% of SMGs contain an AGN (Swinbank et al., 2004), a higher fraction than is seen in local ULIRGs, based on rest-frame H$\alpha$ spectroscopy. We know from more local observations that sufficiently obscured AGN need not be apparent in broad H$\alpha$, and that the AGN fraction in ULIRGs is luminosity dependent (eg. Farrah et al., 2003) so these two results are not in contradiction. The energetic importance of AGN in these systems, though, remains unclear. Alexander et al. (2005) conclude that starbursts, rather than AGN, are the dominant power source in most of their objects. However, Alexander suggests that the X-ray/far-IR ratio in SMGs is less than that found for typical QSOs, and that at least some SMGs are largely AGN powered. Furthermore, AGN dominated sources with Compton thick obscuration (N$_H > 10^{24} cm^{-2}$) (eg. Iwasawa et al., 2005) may also exist among the SMG population, similar to `Quasar 2's seen in the SWIRE survey (Polletta et al., 2006). Direct detection of the AGN emission in such objects will be very difficult. However, the energetic contribution of any obscured AGN can be assessed by examining broad-band spectral energy distributions (SEDs) in the rest-frame near-, mid- and far-IR (eg. Farrah et al., 2003). 

The derivation of the far-IR luminosity in these objects is also an issue which can be addressed by broad band SED studies. The emitted-frame far-IR emission peak lies between 60 and 150$\mu$m for all reasonable dust temperatures ($\sim$50-20K). Even at z=2.5, this peak is only just accessible to ground-based observations for the coldest feasible dust. The determination of the dust temperature in SMGs is thus very difficult, and yet the derived SMG luminosity is a strong power of temperature. Thus if dust temperatures are uncertain to just a factor of 2, the SMG luminosity can be uncertain by more than an order of magnitude. The addition of data points, or simply flux limits, at 70 and 160$\mu$m in the observed frame will provide useful new constraints to the underlying dust temperature and start to solve this problem. Pope et al. (2006), for example, use Spitzer data in the GOODS-N region combined with the GOODS-N 'supermap' to conclude that submm sources on average have a cooler dust temperature than local ULIRGs. Similar conclusions are reached by Kovacs et al. (2006) and Chapman et al. (2005). This leads to an over-estimate of submm source bolometric luminosities if a local ULIRG SED is assumed. Moreover, Efstathiou \& Rowan-Robinson (2003) have suggested that some SMGs have a much cooler dust temperature than would be the case if they were simply high-redshift ULIRGs. These systems would be better fitted by a cool (T$\sim$20--30K) cirrus-like SED, implying a lower redshift and less extreme star formation rate and far-IR luminosity than would otherwise be thought. The overall range of galaxy properties in the SMG population is also still unclear. While the bulk appear to lie at z$\sim$2.5, as found by the spectroscopic followup of radio-identified SMGs by Chapman et al. (2005), even in that work several much lower redshift sources were found (eg. SMM J030226.17+000624.5, aka. CUDSS 3.8, which has a spectroscopic redshift of 0.088 (Clements et al., 2004)). Some of these low redshift objects may in fact be foreground gravitational lenses of background SMGs (Chapman et al., 2002). An examination of the optical-to-far-IR SED of such sources would allow us to say whether a lensed background source is likely, or if the observed submm emission could be accounted for by the presence of an unexpectedly large mass of cold, cirrus-type, dust at low redshift.

Despite much detailed study, the overall stellar population in SMGs remains difficult to trace since the far-IR luminosity is dominated by high mass young stars. Knowledge of this is important since it will indicate the evolutionary status of these systems. This also relates to the initial mass function (IMF) of the underlying starburst. Baugh et al. (2005) have suggested that starbursts SMGs have an IMF skewed to high mass so that the overall stellar mass will be less than otherwise expected. This is the only way in which the SMGs can be accounted for in their semi-analytic models of galaxy formation. Since SMGs are far away and intrinsically optically faint, the presence of lower mass stars is poorly constrained by optical and near-IR observations. The flux received in the optical corresponds to the rest-frame UV and will also be dominated by young stars. The 1.6$\mu$m peak in the SED of normal galaxies due to the bulk of the stellar population (Sawicki, 2002) is redshifted to 5.6$\mu$m at z$\sim$ 2.5 (Simpson \& Eisenhardt, 1999). Observations in the mid-IR, from 3 to 8 $\mu$m are thus ideally suited to placing constraints on the overall stellar mass of these systems (see eg. Borys et al., 2005).

The Spitzer Space Observatory, with instruments operating from 3.6 to 160$\mu$m (Werner et al., 2004) can provide the necessary mid-to-far-IR fluxes required to fill the SED gap between observed frame wavelengths of 2$\mu$m and 850$\mu$m required by the studies described above. SED fitting (eg. Rowan-Robinson et al, 2003) over this broad wavelength range can determine the role of obscured AGN (Farrah et al., 2003), examine the nature of the dust emission, and examine the bulk of the stellar population. These same SED fitting methods are also capable of providing good photometric redshift estimates (Rowan-Robinson et al., 2005; Babbedge et al., 2006), especially if deep optical and/or near-IR data are also available. Previous attempts at photometric redshift estimation for SMGs have set only weak constraints on redshifts either using optical and near-IR data (eg. Clements et al., 2004) or using a combination of radio and submm data (eg. Aretxaga et al., 2007). The addition of Spitzer data allows for much more accurate redshift estimation, possibly as good as $\Delta z / (1+z) \sim$ 0.04 (Rowan-Robinson et al 2007). The efforts necessary to acquire spectroscopic redshifts for large samples of SMGs (e.g. Chapman et al. 2005) might thus be avoided.

Spitzer studies of SCUBA sources to date have generally been based on relatively small samples of sources which are often, as here, pre-selected to have radio associations. Ivison et al. (2004) looked at a sample of 9 MAMBO selected sources (1.2mm) of which 4 were also SCUBA sources, while Frayer et al (2004) examined radio sources in the Spitzer First Look Survey of which 7 had $> 3\sigma$ detections with SCUBA. Egami et al. (2004) similarly looked at radio selected sources, this time in the Lockman Hole East GTO survey, of which 7 were SCUBA sources. In all cases the SCUBA sources were found to have properties similar to those expected of z=0.5---4 far-IR luminous galaxies. Mid-IR colours were used by Ivison et al. and Frayer et al. to identify AGN powered sources, indicating that about 75\% of the SCUBA sources are starburst powered, a result confirmed by IRS mid-IR spectroscopy of smaller samples (eg. Valiante et al., 2007). An interesting aspect of the Ivison et al. work is that nearly all of the MAMBO/SCUBA sources are identified at 24$\mu$m at flux limits comparable to those of the SWIRE survey discussed here. The largest Spitzer study of submm sources to date is that of Pope et al. (2006) using the GOODS-N Spitzer data and the GOODS-N submm 'super-map', which combines photometry, jiggle-map and scan-map submm observations into a single large field (165 arcmin$^2$) (Borys et al., 2003). Of the 35 submm sources in the 'super-map' 21 have secure Spitzer identifications with plausible counterparts for another 12. An alternative, statistical approach to examining the submm properties of Spitzer sources was conducted by Serjeant et al. (2004), who stacked SCUBA data from the 8mJy survey (Scott et al., 2002) at the positions of Spitzer sources in the Early Release Observations (Egami et al., 2004; Huang et al., 2004) to produce average submm properties for various classes of Spitzer object. Statistical detections are made for 5.8 and 8 $\mu$m, and marginally for 24$\mu$m sources. Dye et al. (2006) produce similar results for objects in the CUDSS survey. In summary, our understanding of the properties of submm sources in the Spitzer wavebands is at a relatively early stage. These results so far confirm that the submm population is made up of objects not too dissimilar from local ULIRGs.

The rest of the paper is structured as follows. In Section 2 we provide  a brief summary of the survey data that is the basis for the current paper. Section 3 describes our results for identifying SHADES/radio sources with SWIRE sources, including notes on individual sources of interest. Section 4 describes the methods we use to estimate redshifts and determine the nature of the source SEDs, and presents the results for this analysis. In section 5 we discuss the implications of these results, and we draw our conclusions in section 6. We assume the standard `concordance model' cosmology throughout ie. $H_0=70 {\rm km\,s^{-1}\,Mpc^{-1}}$, $\Omega_{M}=0.3$, $\Omega_{\Lambda}=0.7$.

\section{The SHADES, SWIRE and SXDF Surveys}

The SHADES survey (Mortier et al. 2005; Coppin et al. 2006) is the largest deep, blank field submillimetre survey conducted to date. The final survey was designed to cover two 0.25 sq. deg. fields to a 3$\sigma$ sensitivity of 8mJy using the SCUBA instrument at JCMT in grade 2---3 weather ($\tau_{CSO} = 0.05-0.1$).The two survey regions are in the Lockman Hole region and in the Subaru-XMM deep field (SXDF) region. The demise of SCUBA in 2005 left the survey only about 40\% complete, but this is still larger than any previous submm survey. The current paper focuses on the SXDF region, centred at 02 18 00, $-$05 00 00 (J2000), and covering 247.9 arcmin$^2$ to the required sensitivity. Within this region, 60 sources have been uncovered (Coppin et al. 2006). Data reduction, map making, calibration and source extraction are detailed in Coppin et al. (2006), which also contains more details on the SHADES survey itself.

Radio identifications within the SHADES survey regions are presented by Ivison et al. (2007). The radio data were obtained using the National Radio Astronomy Observatory's (NRAO\footnote{NRAO is operated by Associated Universities Inc., under a cooperative agreement with the National Science Foundation.}) Very Large Array (VLA) in its A, B and C configurations, yielding 1.4-GHz images with resolutions of 1.7\,arcsec {\sc fwhm} and noise levels of 8.4\,$\mu$Jy\,beam$^{-1}$ in the SXDF SHADES field. The techniques used for data reduction and analysis are described in detail by Biggs \& Ivison (2006). Of the 60 SMGs uncovered by SCUBA in this region, 37 have one (or more) radio identifications in this data.

The SWIRE survey is the largest of the Spitzer Legacy surveys (Lonsdale et al. 2004). It covers a total of 49 sq. deg. in all seven of the continuum bands available to Spitzer ie. 3.6, 4.5, 5.8, 8$\mu$m using the IRAC instrument (Fazio et al., 2004) (referred to as IRAC bands 1, 2, 3 and 4 respectively) and 24, 70 and 160 $\mu$m using the MIPS photometer (Rieke et al., 2004) (referred to as MIPS 1, 2 and 3 respectively). The (5 $\sigma$) sensitivities achieved in these bands are typically 3.7, 5.3, 48, 37.7 $\mu$Jy in the IRAC bands, and 230$\mu$Jy in the MIPS1 (24$\mu$m) band and 20 and 120 mJy at 70 and 160$\mu$m. Catalogues for SWIRE sources are produced by source extraction at each SWIRE wavelength. These are later band-merged into a unified catalogue providing fluxes for each source
at all wavelengths at which it is detected. The SHADES SXDF region is covered by the SWIRE XMM-LSS field. This is centered at RA = 02 21 20, DEC= -04$\deg$ 30' 00" and covers a total area of 9.1 sq. degrees, encompassing the whole of the SHADES region.

The Subaru/\textit{XMM-Newton\/} Deep Field (SXDF) optical observations (Kodama et al., 2004) are among the deepest optical images ever obtained from the ground.Observations were carried out in the $B,V,R,i'$ and $z'$ broad bands and two narrow bands. We here use only the $B,V,R,i'$ and $z'$ bands. The depths reached are 27.5, 27.5, 27, 27 and 26 respectively in the five bands (AB magnitudes), measured in a 2'' diameter aperture (Furusawa et al., in preparation). The survey covers an area of 1.2 sq. deg. centered at 02 18 00 $-$05 00 00 (J2000), encompassing the whole SHADES region, though some small part of this survey is masked out by the effects of bright stars. The observations were taken in good conditions with seeing better than 1". Optical catalogs from this survey are publicly available from \begin{verbatim} http://www.naoj.org/Science/SubaruProject/SDS/ \end{verbatim}, while the SWIRE catalogs can be found at \begin{verbatim} http://swire.ipac.caltech.edu\end{verbatim}.

\section{The SWIRE-SHADES Sources}

The identification of counterparts to SMGs at other wavelengths is made difficult by large submm beam sizes --- $\sim$15" for SCUBA --- (Blake et al., 2006) and is further complicated by the closeness of typical deboosted (Coppin et al., 2006) detected submm fluxes ($>\sim$4mJy for SHADES) to the confusion limit ($\sim$2mJy (Eales et al. 2000; Blain et al. 1998; Hughes et al. 1998) and the inherent faintness of most SMGs in the optical and near-IR bands (see eg. Clements et al., 2004). Nevertheless, deep radio observations, reaching a 1$\sigma$ sensitivity $\sim 10\mu$Jy at 1.4 GHz, (Ivison et al. 2007) have been successful in identifying a substantial fraction ($\sim$65\%) of the SMG population and, by providing positions for the radio counterparts of SMGs accurate to $\sim$0.5", allowing SMG counterparts at other wavelengths to be identified and followed up (see eg. Chapman et al. 2005). Frayer et al. (2004) and Ivison et al. (2004) have also found that many SMGs are also detected in deep 24$\mu$m Spitzer observations, providing a complementary route to finding counterparts to SMGs.

In this paper we examine the optical-to-submm SEDs of SMGs. The first step towards this goal is to find counterparts to the SMGs with greater positional accuracy than can be achieved in the submm alone. Once localized in this way, counterparts to each SMG in the optical and IRAC bands can be determined and the SED extracted. This procedure goes through the intermediate step of a radio (or 24$\mu$m) identification since the high source densities in the optical and IRAC bands at the fluxes needed to detect an SMG make it difficult to acquire an unambiguous identification in these bands themselves.

There are 60 SMGs in the SHADES survey of the SXDF. Of these, 38 have good radio associations (P$<$0.05, where P is the probability of a chance association).  21 of the 60 sources have good (P$<$0.05) 24$\mu$m associations. In all but one case (SHADES-SXDF 77) these are sources with good radio identifications as well. Some of the radio and 24$\mu$m associations are multiple. We thus have a total of 39 plausibly identified SMGs, 38 from radio, 1 from 24$\mu$m. The reliabilities of these P values were investigated through extensive simulations by Ivison et al. (2007). They conclude that the values are reliable, but that the number of multiple identifications is higher than expected on the basis of random radio positions. They ascribe these multiple identifications to genuine physical associations though this does not necessarily mean that more than one radio source is responsible for the radio emission. This is discussed further below (Section 5.5). They further conclude that it is unlikely that any of our sources are the result of confusion ie. the mixing of two physically unrelated SMGs in the same SCUIBA beam. For more details see Ivison et al. (2007).

We conduct a search in the SWIRE catalog for infrared sources at the radio positions  (and for SHADES-SXDF 77 the 24$\mu$m position). We find SWIRE associations for 32 of these sources (31 from radio plus SHADES-SXDF 77 from 24$\mu$m), an identification rate of 82\% for the radio associated SHADES sources, and 53\% overall. The results of this search are shown in the appendix, which presents postage stamp images in $i$, IRAC 1 (3.6 $\mu$m), and MIPS 1 (24$\mu$m) for each identified source, with the SWIRE, radio and SCUBA positions indicated. In all but two cases (SHADES-SXDF69 and 88) the P statistics for these sources, indicating the probability of a chance association between the submm, radio and/or SWIRE source, are $<$0.05. For more details of the P value calculations see Ivison et al., (2007). SHADES-SXDF69 and 88 have P values $<$0.1, but their identifications are detected at both radio and 24$\mu$m which adds credence to the association, as well as good SED fits. However these identifications should probably be treated with some caution. Two sources (SHADES-SXDF 14 and 24) have ambiguous radio identifications where one radio source with a good P value is not associated with any optical or SWIRE source. We here examine the other candidate radio identifications and find that they produce reasonable SED fits. Nevertheless these identifications are not secure. One further source, SHADES-SXDF74, has its radio ID offset from the supposed optical/SWIRE counterpart, and so should also be treated with caution.

Once a SHADES source has been associated with a SWIRE and radio source, we also extract its optical fluxes from the SXDF optical data. One additional source (SHADES-SXDF 12) is found to have an optical counterpart at the radio position, but no SWIRE infrared counterpart. The results of these associations are presented in Table 1. We only use the IRAC bands and MIPS1 (24$\mu$m) channels to look for associations in the SWIRE catalog since the sensitivity and angular resolution in MIPS2 and 3 (70 and 160$\mu$m) are much poorer than at shorter wavelengths. However, for those sources identified at shorter wavelengths, we examine the catalogs for 70 and 160$\mu$m fluxes associated with them, finding five sources with useful fluxes. These are listed in Table 2. The sources detected by MIPS 2 and/or 3 all turn out to be either at lower redshfits (SHADES-SXDF 21, 77 and 77) or to be amongst the most luminous in the sample (SHADES-SXDF 5, 28).

The catalog used for 24$\mu$m associations was based on a deeper extraction of the SWIRE data than in the current SWIRE public release. We used different SExtractor detection thresholds producing a 2.5$\sigma$ catalog rather than the publicly available 5$\sigma$ catalog, allowing us to go deeper.

Seven sources with radio associations do not have SWIRE associations. Of these, six (SHADES-SXDF 16, 18, 38, 40, 50 and 55) do not have optical associations either. The remaining one (SHADES-SXDF 12) has optical but no SWIRE counterparts. Its optical fluxes have been extracted from the SXDF catalog and are given in Table 1. The six optical non-detections clearly have a very large submm/optical flux ratio, and must be considered potential high redshift (z $\geq$2.5) sources on this basis. This is largely born out by the radio-submm-far-IR photo-z estimates for these sources from Aretxaga et al. (2007), where most are expected to lie at z$\ge$2.5.

Of the 31 SHADES sources with plausible radio identifications and matching radio/SWIRE data, the one 24$\mu$m and non-radio identified source, and the one radio and optical source without matching SWIRE fluxes (33 sources in total), we confirm that 32 have believable optical-to-far-IR SEDs. For one source (SHADES-SXDF 31) the plausible radio identification does not produce a reasonable SED when the submm flux from SCUBA is combined with the optical-to-mid-IR fluxes obtained by SWIRE. Instead a nearby SWIRE/optical 
source that does not have observed radio emission appears to be a better SED match to the submm flux.
This suggests that SMG identification through their radio counterparts is quite reliable since only 1/32
(3\%) radio identified sources with SWIRE and/or optical data is not confirmed through SED analysis.

The one 24$\mu$m only identification (71) provides a reasonable SED as does the one optical/radio identification without a SWIRE detection (12).

Of our final list of 33 associations between SHADES sources and SWIRE and/or SXDF sources, 31 have associations at 3.6$\mu$m, 29 at 4.5$\mu$m, 11 at 5.8$\mu$m, 12 at 8$\mu$m, 27 at 24$\mu$m, 4 at 70$\mu$m and 3 at 160$\mu$m. One source has an optical and radio association but no SWIRE fluxes.

\begin{table*}
\vbox to220mm{\vfil Landscape data table available from author D.L. Clements
  \caption{}
 \vfil}
 \label{landfig}
\end{table*}

\begin{table}
\begin{center}
\begin{tabular}{ccc} \hline
SHADES-SXDF\#&F70 (mJy)&F160 (mJy)\\ \hline
5&19.2 $\pm$ 3.2&128.7$\pm$15.1\\
21&121.2$\pm$23.8&301.4$\pm$30.8\\
28&18.5$\pm$3.1&-1\\
77&-1&103.5$\pm$15.4\\
119&7.3$\pm$3.2&-1\\
\end{tabular}
\end{center}
\caption{Sources detected at 70 and/or 160$\mu$m}
-1 indicates a non-detection, uncertainties are 1$\sigma$.
\end{table}

\subsection{Ambiguous Identifications}

There are a number of sources which have ambiguous identifications. These include SHADES objects where there are more than one possible associated radio source (eg. SHADES-SXDF 52) and others where, as will be shown below, the photo-z and SED fitting for the candidate counterpart identified by the radio data turns out to have problems and where there are other, alternative SWIRE sources that might be associated with the submm flux. The approach we have adopted is to 
use a combined $\chi^2$ for the optical and 3.6-850 $\mu$m infrared template fitting to select the optimum photometric redshift (where
there are aliases) and association (where these are ambiguous). See the Appendix for discussion of individual sources.

\section{The SEDs of SHADES-SWIRE sources}

\subsection{Photometric Redshifts and Galaxy Classificattions}

With the extraction of optical-to-far-IR fluxes for each SMG, we can apply photometric redshift techniques to this data to obtain both a redshift estimate and a classification for the object's properties in both the rest-frame UV-optical-IR and mid-to-far-IR. The technique we apply is I{\scriptsize MP}Z (Rowan-Robinson 2003, Babbedge et al. 2004,
Rowan-Robinson et al 2007).

Our I{\scriptsize MP}Z  analysis is a two stage process whereby we first obtain a redshift and classification in the optical and IRAC 1 \& 2 bands, and then fit the remaining emission in IRAC, MIPS and SCUBA bands, together with any IR excess in the IRAC 1 and 2 bands, using far-IR templates.

\subsubsection{Optical/IRAC Photo-z Estimation}

I{\scriptsize MP}Z uses seven galaxy templates (E, protoE, Sab, Sbc, Scd, Sdm and Starburst - for details of these templates, including star formation history, extinction etc., see Rowan-Robinson et al., 2007) and three (type 1) AGN templates from Babbedge et al. (2004), with the additional proto-elliptical from Maraston (2005), with IGM treatment and Galactic extinction corrections. At redshifts $>$2.5 the E template is dropped since old ellipticals are unlikely to be found at such high redshifts. Variable extinction, in addition to that already included in the templates (see Rowan-Robinson et al. (2007) for details) is allowed in the fits and combined with the templates.  Additional extinction, A$_{\rm V}$, limits of 0 to 1 are used for galaxies, and A$_v\leq0.3$ for AGN, unless there is clear evidence for a dust torus, in which case $A_V$ is allowed to be up to 1.  A prior expectation that the probability of a given value of A$_{\rm V}$ being `correct' declines as $|$A$_{\rm V}|$ moves away from 0 is applied.  This is introduced by minimising $\chi^{2}_{red}$ + $\alpha$A$_{\rm V}^2$ rather than $\chi^2$ (where $\alpha$=1.5).
Rowan-Robinson et al. (2003b) found it necessary to apply absolute magnitude limits to exclude unlikely solutions, such as super-luminous sources at high redshift or very underluminous objects at low redshift.  Here, the following limits are used:  Absolute magnitude limits of ( -17 +z)$<$M$_B<$[-22.5+z]  for $z<$2.5, and -19.5$<$M$_B<$-25.0 for $z>2.5$ for galaxies and -21.7$<$M$_B<$-27.7 for AGN.
The full justification of the $A_V$ and $M_B$ priors is given in Rowan-Robinson et al (2007).  Although a very 
small minority of SWIRE sources ($<1\%$) may get better fits with extra $A_V > 1$, inclusion of this possibility leads to
increased outliers due to aliasing.  We have considered the possibility of higher extinctions in our SED fitting, but
it can be seen from Table 2 that the submillimetre galaxies do require high optical extinction beyond what is already in the templates. 
The resulting photometric redshifts are expected to be accurate to $\sim0.04$ in $(1+z)$, based on previous studies (Rowan-Robinson 2003; Babbedge et al 2004; Rowan-Robinson et al. 2007).  
Due to aliasing and photometric variability issues  the accuracy for the AGN will be lower, $\sim0.2$ in $(1+z)$.  Franceschini et al. (2005) also found lower photometric redshift accuracy for their AGN in their SWIRE/$Chandra$ study, whilst Kitsionas et al. (2005) found that, for their X-ray selected XMM-$Newton$/2dF sample, the photometric redshift accuracy for AGN was $<0.2$ in $\Delta z/(1+z)$ for 75 per cent of the sample.  Here, in order to successfully deal with cases where there is significant dust torus contribution to emission at 3.6 and 4.5$\mu$m, a `double pass' of the catalogue through I{\scriptsize MP}Z is carried out:  In the first pass, the 3.6 and 4.5$\mu$m data are not included in the fit if S(3.6)/S(r$\arcmin)>$3; in the second pass the 3.6 and 4.5$\mu$m data are included in the fit provided S(3.6)/S(r$\arcmin)<$300, except when the mid-IR fitting resulted in an AGN dust torus fit.  Any mid-IR excess is fitted separately using dust templates in the next stage.

\subsubsection{Mid-IR to Submm SED fitting}

This follows the technique of Rowan-Robinson et al. (2005).   A source that has obtained a best-fitting template and photometric redshift, based on its B band to 4.5$\mu$m detections as set out above, now has its IR excess calculated by subtracting the galaxy model fit prediction from the 4.5 to 850$\mu$m data.  At least two of these bands need to exhibit an IR excess (one of which is required to be 8 or 24$\mu$m) for a dust template to be fitted. All of the SHADES sources exhibit such an excess.
This excess is then characterised by finding the best-fitting out of cirrus dust (Efstathiou \& Rowan-Robinson, 2003), M82 or Arp220 starbursts (Efstathiou et al., 2000) or AGN dust torus IR templates (Rowan-Robinson, 1995).  Each template is the result of radiative transfer models. While it is possible that for {\em individual} galaxies one could obtain better {\em individual} fits, we here wish to characterise only the broad types of IR population.
Sources are allowed to be fitted by a mixture of an M82 starburst and cirrus, or by a mixture of an M82 or Arp220 starburst
and an AGN dust torus, since it was found that, often, both components were required to represent the IR excess.  This mirrors the conclusions of Rowan-Robinson \& Crawford (1989) from fitting mid-IR SEDs to IRAS sources, and the findings of Rowan-Robinson et al. (2005) for Spitzer-SWIRE.  In the case of a single band IR excess, an M82 SED is assumed.

Where there is more than one possible association with the 850 $\mu$m source, or where the photometric redshift solution shows the presence of aliases, we have combined the optical and infrared $\chi^2$s to select the best solution.
The power of the 850 $\mu$m flux in selecting between infrared template fits, and its insensitivity to redshift, makes
this an effective procedure.
The results of the photo-z and mid-IR SED fitting are summarized in Table 3.

\section{Discussion}

\subsection{The Redshift Distribution of SHADES sources}

The redshift distribution of the sources discussed here is shown in Figure 1, compared to the spectroscopic redshift distribution for radio-selected/identified submm sources found by Chapman et al. (2005). The median photometric redshift for our SWIRE-identified sources is z = 1.44. This compares to the median of z = 2.2 from the spectroscopic and photometric redshifts for 33 Spitzer-identified SMGs in the GOODS-N 'super-map' radio-detected subsample (Pope et al, 2006), (the 21 secure identifications in the 'super-map' sample have the same median redshift), a median of 2.15 for the objects with spectroscopic redshifts from Chapman et al. (2005) and a median redshift of 1.52 for the Lockman Hole SHADES sources (Dye et al., 2007). Once account is taken for the effects of the spectroscopic redshift desert at around z$\sim$1.5, the median redshift from Chapman et al. falls to z$\sim 2$.

There thus appears to be a lack of high redshift sources in our sample compared to Chapman et al. and Pope et al. though we are in reasonable agreement with Dye et al.. The radio flux limit of our identifications is as deep as all but two of the Chapman fields (Lockman and SSA-13) which are a factor of $\sim$2 deeper,  but which account for only 13 of their 76 redshifts. The redshfts measured in these two fields are somewhat higher than the rest of the sample. The GOODS-N field has a comparable depth to the deepest Chapman fields, ie. 1$\sigma \sim 5\mu Jy$, so this could account for the difference between the current paper and the Pope et al. (2006) redshift distribution.

The seven radio-identified sources in our sample without SWIRE/optical counterparts are potentially at high redshift or might be very heavily obscured. If we combine the redshift estimates for these sources from their radio-submm-far-IR properties from Aretxaga et al. (2007), using their $z_{phot}^{MC}$ value, then the median redshift for our sample rises to 1.9. However, there are large uncertainties in the $z_{phot}^{MC}$ estimates, with 90\% confidence intervals typically covering the range 2$\le$z$\le$3, and sometimes much larger. The radio non-detected sources may lie at still higher redshift. The radio non-detected subsample in  GOODS-N has a median z = 2.3, for example (Pope et al., 2005). The GOODS-N Spitzer data is significantly deeper than the SWIRE data, with, for example, 24$\mu$m fluxes reaching a 5$\sigma$ sensitivity of 24$\mu$Jy compared to the SWIRE limit of 106$\mu$Jy, allowing their Spitzer identifications to reach higher redshifts.

A comparison between our redshift estimates, based on optical-to-submm fluxes, with those derived from the radio-to-far-IR fluxes by Aretxaga et al. (2007) is shown in Figure 2. As can be seen there is broad agreement between the two results, but there are a small number of sources where things may have gone wrong. Examining this in detail we find that there are six sources (SWIRE-SXDF 1, 3, 12, 30, 74 and 119) whose optical-submm photometric redshift lies outside the 90\% confidence interval given for the radio-far-IR derived redshift from Aretxaga et al. For our sample size we would expect 3 sources to be outside this confidence interval. On examining the SED fits for these discordant objects we find that 4 of them (1, 3, 74 and 119) have cirrus-type dust SED fits. These sources have cooler dust ($\sim$20K) than is normally expected for far-IR-luminous objects and it is likely this factor which is leading the radio-far-IR method to place them at a higher redshift than the method we have used. These sources fully account for the apparent disagreement between our and Aretxaga's redshift estimates. It is also worth noting that the identification for source 74 is potentially unreliable, though there are no indications of this from the photo-z or SED fit. Disagreements at higher redshift may well result form higher luminosity objects having a higher temperature than the ensemble average used by Aretxaga et al (2007), though these disagreements are largely within the acknowledged statistical uncertainties.

In conclusion, the redshift distribution we recover for the SXDF region of SHADES, is consistent with previous work, given the advantages and limitations of the techniques we are using, with the bulk of submm galaxies lying at z$\sim$1.5 -- 2.5. Furthermore, Aretxaga et al. (2007), using radio-submm-far-IR techniques, conclude that the SXDF SHADES field has a somewhat lower redshift peak than the Lockman SHADES field (2.2 compared to 2.6). The results we derive for the radio/SWIRE identified sources are consistent with this conclusion.


\begin{figure}
\epsfig{file=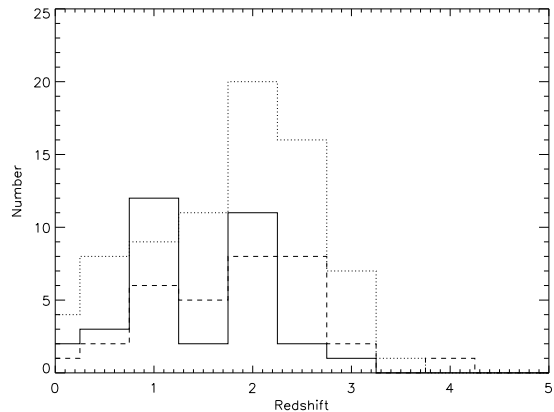,height=8cm, angle=90}
\caption{Redshift distribution of identified SCUBA/SHADES sources. 
The solid line shows a histogram of the redshift distribution of the identified sources derived using ImpZ. The dotted line shows the redshift distribution from Chapman et al. (2005) derived from spectroscopic observations of radio identified and/or selected submm sources in a variety of other submm surveys, while the dashed line shows the redshift distribution for the Pope et al. (2006) GOODS-N 'supermap' submm galaxies, including both spectroscopic (16, from Chapman et al., (2005)) and photometric (9 with optical and Spitzer data, 8 purely from Spitzer data) redshifts for both solid (21) and tentative (12) identifications.}
\end{figure}

\begin{figure}
\epsfig{file=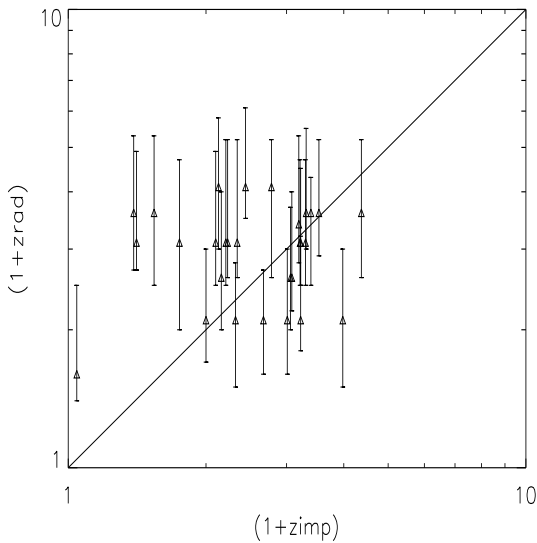, height=8cm, width=8cm, angle=90}
\caption{Comparison of current redshift estimations and those using radio-far-IR Fluxes. 
Comparison between the photometric redshifts obtained here (zimp) with those obtained by Artexaga et al. (2007) using a radio-far-IR based method (zrad). The error bars shown are 1$\sigma$ uncertainties. We use their $z_{phot}^{A}$ parameter for this plot.}
\end{figure}

\begin{figure*}
\epsfig{file=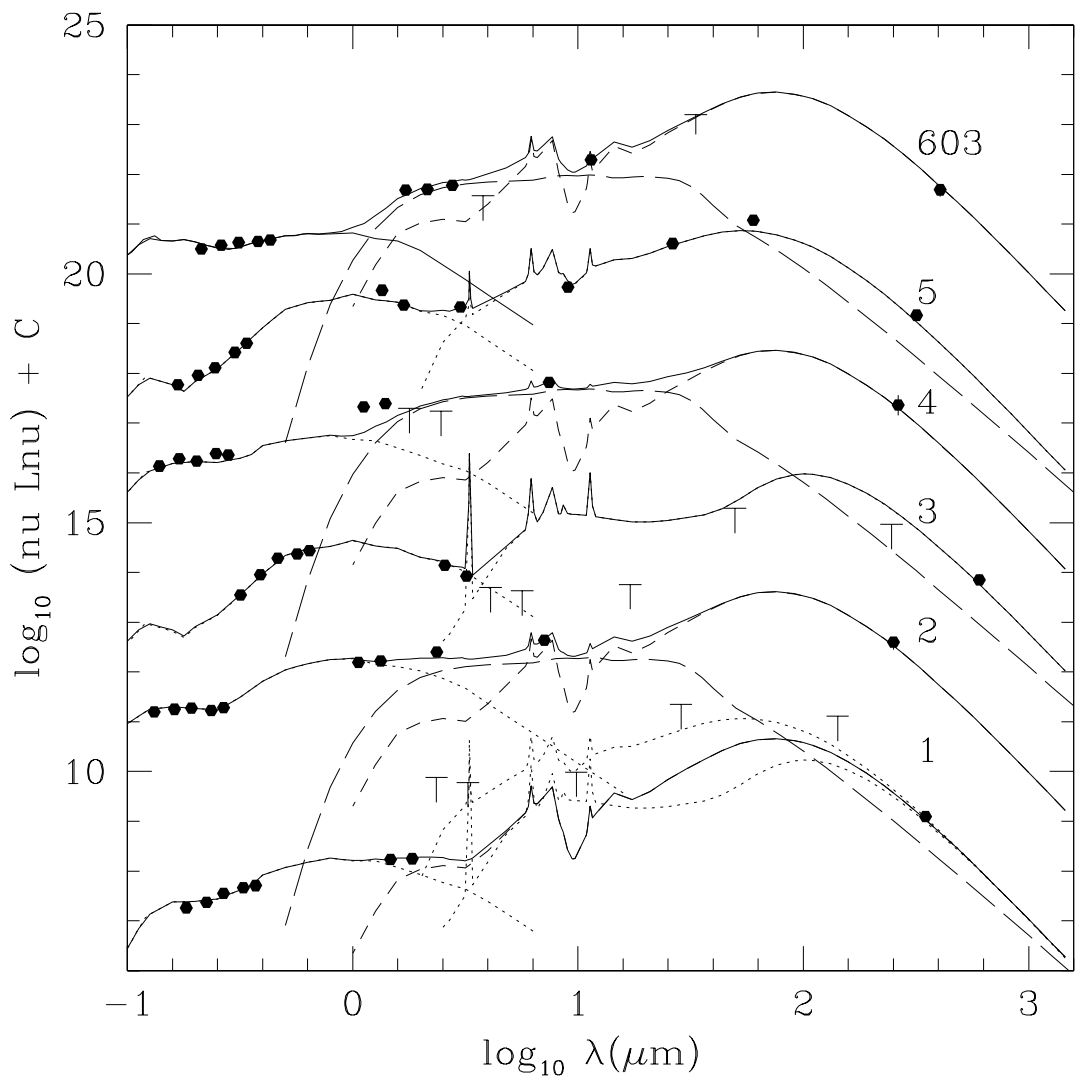, width=15cm}
\caption{SED fits to SHADES-SWIRE sources.
All of the 33 associations were succesfully modelled with the IMPz optical template fitting code (Babbedge et al 2004; Rowan-Robinson et al., 2007), generating photometric redshifts.  The infrared and submm excesses with respect to the starlight/AGN fits were modelled with a set of four templates: cirrus,
M82 and Arp 220 starbursts, and an AGN dust torus model, as described
by Rowan-Robinson et al (2005). See Table 2 for details of the fits. The resulting SEDs are plotted here. Black dots are data points, error bars for these are smaller than the dots. Crosses indicate flux upper limits if these have proved significant to the fits. The fitted optical-to-far-IR SED is indicated by a solid line. Where two components contribute to the fit, typically a starburst and an AGN, these are also shown separately. AGN torus with long dashed lines, M82 starburst with dotted lines and Arp220 starburst with short dashed lines. Where there is a significant infrared excess in the IRAC 1 \& 2 bands the contribution of the optical fit is also shown as a dotted line. For source 1, as a demonstration of the fits, the two non-fitting non-AGN SEDs are also shown as dotted lines.}
\end{figure*}

\begin{figure*}
\epsfig{file=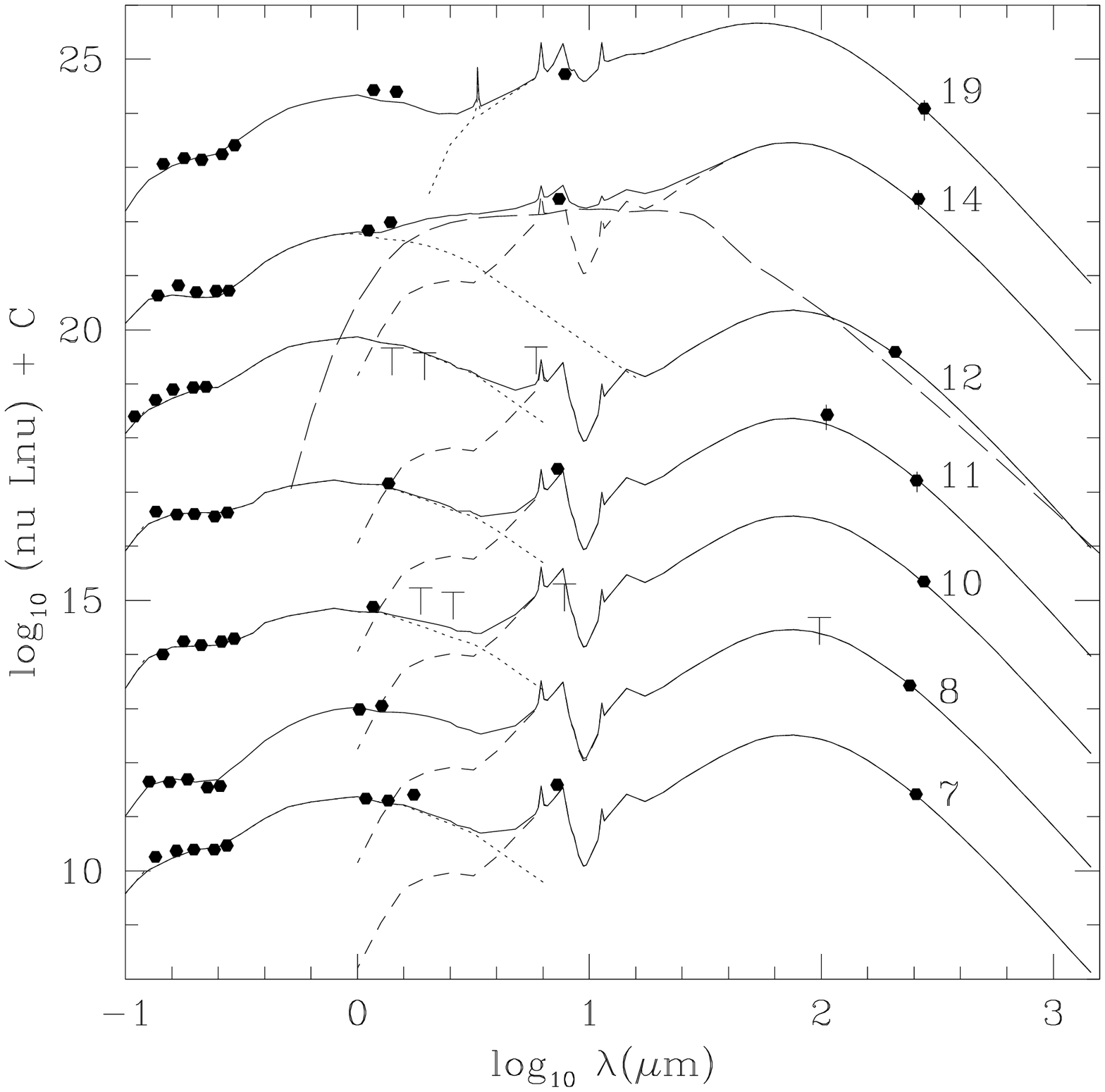, width=15cm}
\contcaption{}
\end{figure*}

\begin{figure*}
\epsfig{file=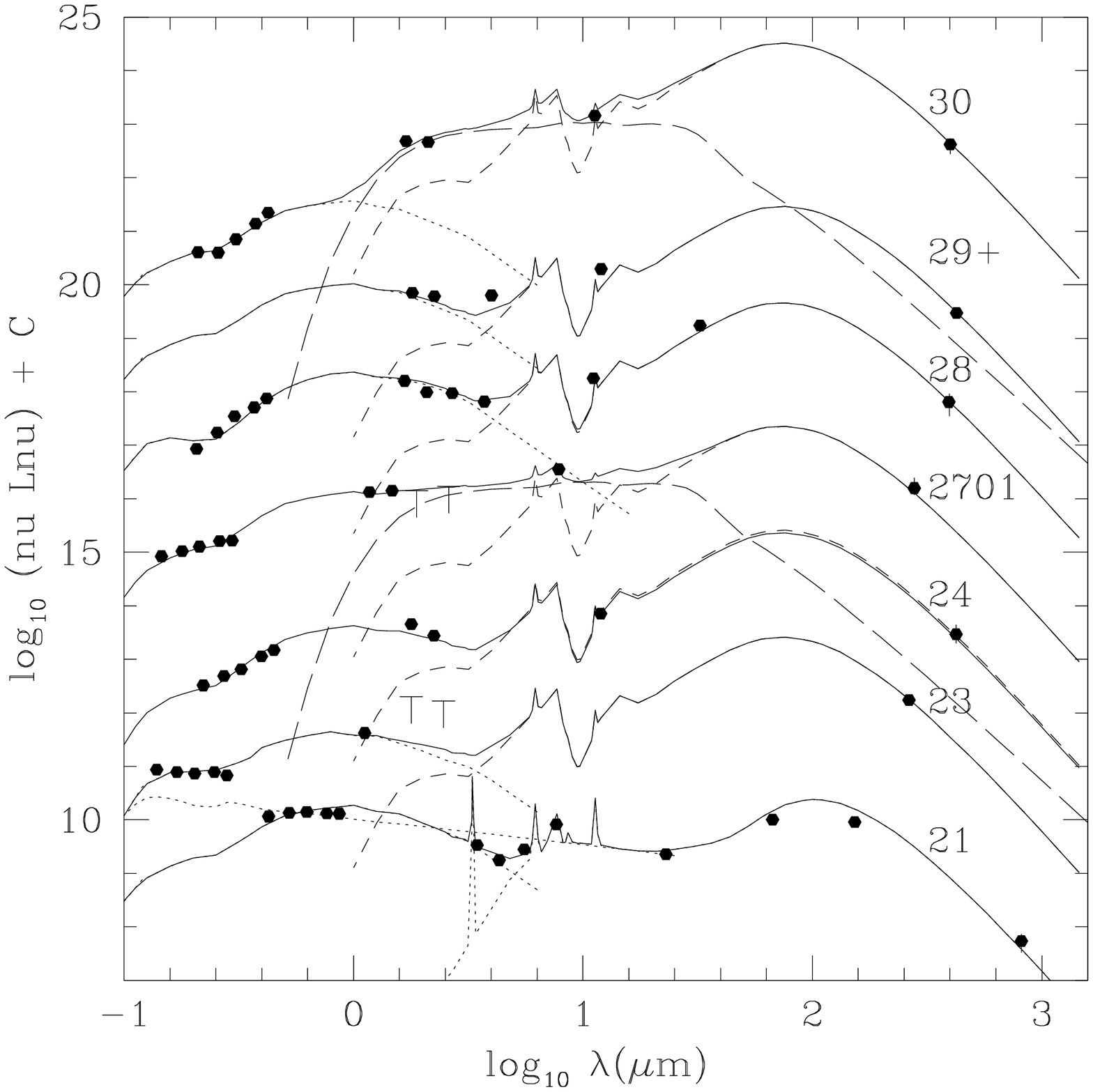, width=15cm}
\contcaption{}
\end{figure*}

\begin{figure*}
\epsfig{file=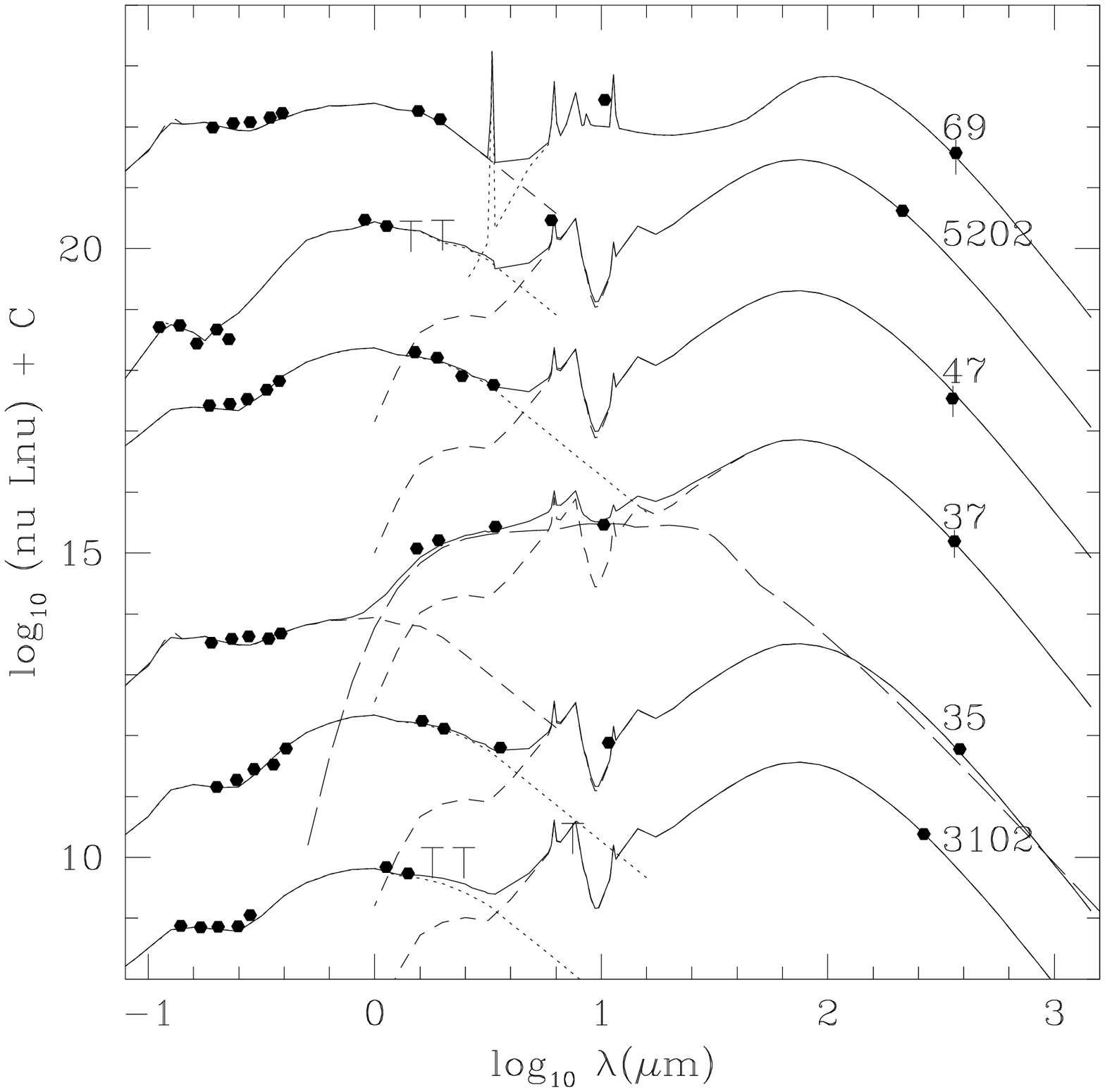, width=15cm}
\contcaption{}
\end{figure*}

\begin{figure*}
\epsfig{file=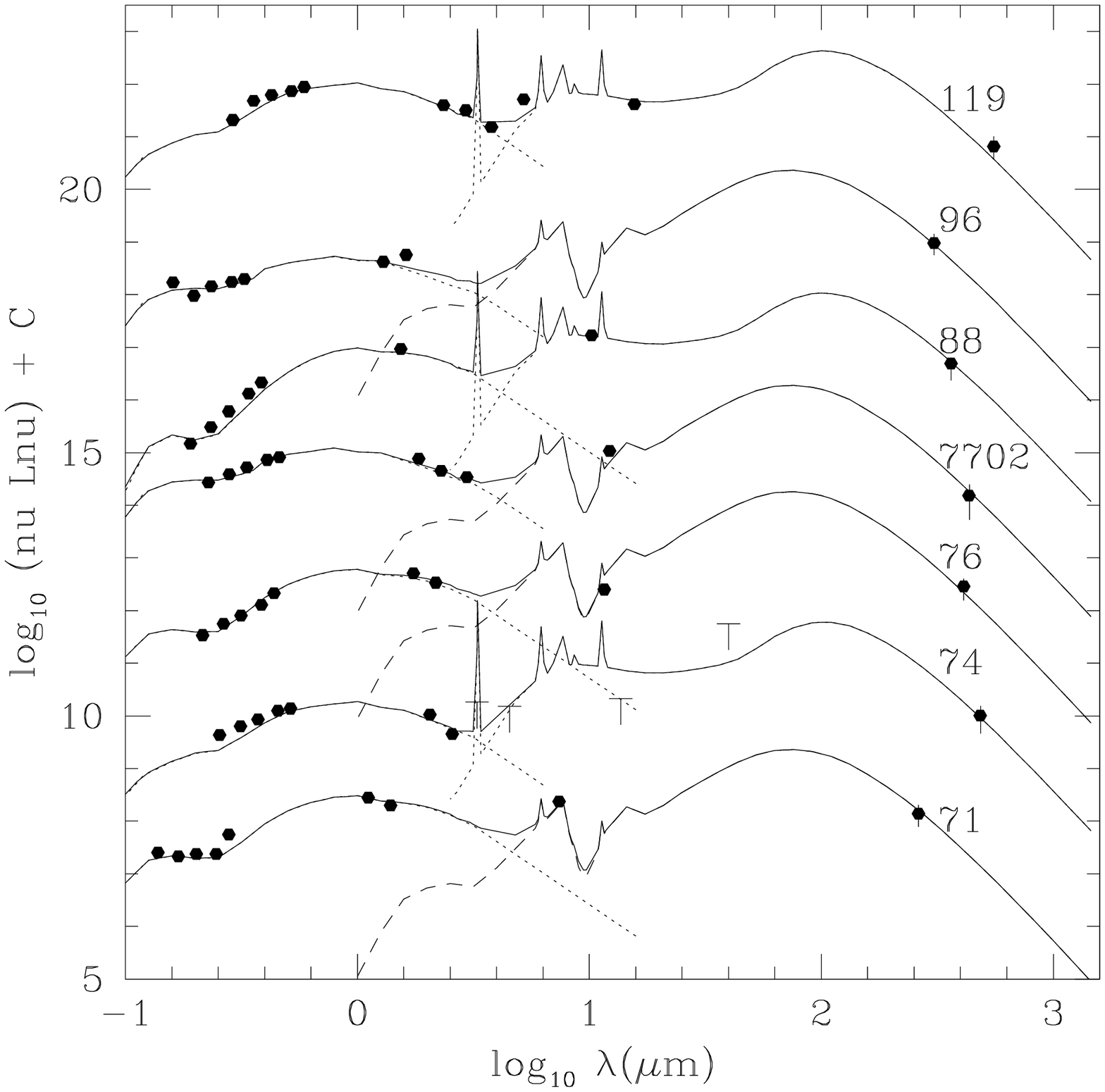, width=15cm}
\contcaption{}
\end{figure*}

\begin{table*}
\begin{tabular}{crrrrrrrrrr}
 & & & & & & &\\
SHADES-SXDF No. & $z_{phot}$ & opt type & $L_{opt}$ & $A_V$ & $A_V^{max}$&C & ir type & $L_{ir}$  & $\chi^2_{opt}$ & $\chi^2_{ir}$\\
&&&&&&&&&\\
1 & 1.44 & Scd & 10.56 & 0.4&3.0 & -2 & 1 & 12.36 & 4.45 & 0.00\\
 & & & & & & &(2 & 13.26) & &\\
 & & & & & & &(3 & 12.76) & &\\
2 & 2.39 & Sab & 11.45 & 0.0 & 0.35& 1 & 3 & 12.75 & 3.38 & 0.13\\
 & & & & & & &4 & 11.60 & & \\
3 & 0.41 & E & 10.56 & 0.0 & 0.0&4 & (1 & 11.96) & 0.63 & - \\
4 & 2.22 & Scd & 11.00 & 0.0&2.6&  6 & 3 & 12.60 & 4.85 & 0.15\\
 & & & & & && 4 & 12.00 & &\\
 5 & 1.67 & Sab & 11.67 & 0.0 &0.35& 8 & 2 & 13.07 & 18.8 & 3.18\\
603 & 1.10 & sb & 10.17 & 0.0 & 4.8&11 & 3 & 10.92 & 0.83 & 0.07\\
 & & & & & & &4 & 11.22 & & \\
7 & 2.31 & Sbc & 11.55 & 0.0 &0.6& 0 & 3 & 12.65 & 1.85 & 0.43\\
8 & 2.53 & Sab & 11.27 & 0.4 &0.75& 2 & 3 & 12.60 & 14.63 & 0.03\\
10 & 2.08 & Scd & 11.13 & 0.2&2.8 & 4 & 3 & 12.68 & 4.18 & -\\
11 & 2.30 & Scd & 11.50 & 0.1&2.7 & 6 & 3 & 12.50 & 4.14 & 0.73\\
12 & 3.07 & Sbc & 12.08 & 0.0&0.6 & 8 & 3 & 12.53 & 3.13 & - \\
14 & 2.24 & Sab & 11.55 & 0.2&0.55 & 11 & 1 & 12.60 & 3.17 & 2.92\\
19 & 2.06 & Sbc & 11.38 & 0.0&0.6 & 13 & 2 & 12.88 & 0.19 & 0.90\\
21 & 0.044 & Sbc & 10.15 & 0.0&0.6 & 0 & 1 & 10.25 & - & -\\
 & 0.044 & QSO & 10.55 & 0.2&0.2 & & & & &\\
23 & 2.22 & Scd & 10.95 & 0.0&2.6 & 1 & 3 & 12.55 & 4.87 & -\\
24 & 1.01 & Sbc & 10.83 & 0.25&0.85 & 3 & 3 & 12.53 & 0.68 & 1.90\\
2701 & 2.06 & Sbc & 11.33 & 0.1&0.7 & 5 & 3 & 12.58& 0.03 & 0.55\\
28 & 1.16 & Sab & 11.52 & 0.2&0.55 & 7 & 3 & 12.72 & 5.66 & 3.90\\
29+ & [1.0] & Sbc & 11.11 & 0.0&0.6 & 9 & 3 & 12.51 & - & -\\
30 & 1.13 & Sbc & 9.67 & 0 &0.6& 11 & 3 & 12.57 & 0.55 & 0.02\\
 & & & & & && 4 & 12.27 & & \\
3102 & 2.21 & Sab & 11.00 & 0&0.35 & -1 & 3 & 12.70 & 2.15 & -\\
35 & 1.23 & Sab & 11.49 & 0.2 &0.55& 1 & 3 & 12.59 & 3.92 & 0.93\\
37 & 1.34 & sb & 9.85 & 0.2 & 5.0&4.5 & 3 & 12.45 & 2.44 & 0.91\\
& & & & & & & 4 & 11.25 & &\\
4702 & 1.39 & Sab & 11.50 & 0.6&0.95 & 7 & 3 & 12.40 & 1.98 & 0.16\\
5202 & 2.98 & E & 11.57 & 0.0 & 0.0&9 & 3 & 12.62 & 3.38 & 3.29\\
69 & 1.31 & sb & 11.79 & 0.2 & 5.0&11 & 1 & 11.96 & 1.37 & 0.66\\
 && & & & & & 4 & 10.05 & &\\
71 & 2.24 & Sab & 11.70 & 0.2&0.55 & -3 & 3 & 12.50 & 3.64 & 0.37\\
74 & 0.75 & Sbc & 10.33 & 0.0 &0.6& 0 & 1 & 11.83 & 0.86 & -\\
76 & 1.07 & Sab & 10.92 & 0.2&0.55 & 2 & 3 & 12.32 & 1.26& 10.18\\
7702 & 0.96 & Scd & 11.28 & 0.1&2.7 & 4 & 3 & 12.34 & 1.51 & 1.28\\
88 & 1.34 & Sab & 11.25 & 0.8&1.15 & 6 & 1 & 12.15 & 5.83 & 0.00\\
96 & 1.78 & Scd & 10.97 & 0.1&2.72 & 8 & 3 & 12.47 & 1.74 & 0.0 \\
119 & 0.54 & Sbc & 11.05 & 0.0&0.6 & 11 & 1 & 11.65 & -&-\\
\end{tabular}
\caption{Properties of SWIRE-SHADES galaxies with plotted SEDs.
Opt. type gives the optical template used for the photo-z fit (Babbedge et al.,2004; Rowan-Robinson et al., 2007);  L$_{opt}$ gives the derived optical B band luminosity; $A_V$ gives the V band extinction additional to that already included in the templates derived from the SED fit; $A_V^{max}$ gives the total extinction to the most obscured component in the template, including both the template $A_V$ and the additional obscuration found by the fitting process (see Rowan-Robinson et al., 2007, for details of extinction to all the components that make up the template SEDs); C gives the offset value used in the SED plots in Fig. 3; IR type gives the fitted far-IR SED template. These are 1: Cirrus; 2: M82 starburst; 3: Arp220; 4: AGN dust torus; L$_{fir}$ gives the derived far-IR luminosity.}


\end{table*}

\subsection{Spectral Energy  Distribution Fits}

As well as estimating redshifts, the photo-z/SED fitting system gives an optical and far-IR SED class that has been fit to the sources. We find that the fitted optical templates are mostly star-forming classes, as would be expected if these sources are high redshift starbursts. Two of the sources, rather surprisingly, are best fitted by an Elliptical or proto-elliptical template in the optical (SHADES-SXDF 3 and 5202). The first of these is a low redshift source (z$_{phot}$ = 0.41) with a cirrus-type far-IR SED. The second E-type source is 5202 which has an Arp220 type far-IR SED on top of an elliptical optical SED at a photometric redshift of 2.98, making it an interesting object for further study. At this high a redshift the E-type template is that of a proto-elliptical from Maraston (2005). It is very faint in the optical, with red optical-to-3.6$\mu$m colours (Davoodi et al., 2006), and flat 3.6-to-4.5$\mu$m colour, indicating both the SED type and the redshift. One further source, 21, has an optical SED that could be fit by a QSO SED in the optical, but this is not consistent with the IRAC 3.6 and 4.5$\mu$m fluxes. Source 21 is an interesting object in that it is  detected in all Spitzer wavebands, both IRAC and MIPS. This is because it is a bright, low redshift dusty galaxy with a cirrus-type far-IR SED.

The far-IR SEDs are more varied than the optical ones, which are dominated by star-forming types. The majority of sources, 23/33, are, as expected, Arp220-like ULIRG SEDs. However, there are also two M82-type starbursts. The M82 SED contains warmer dust, which peaks at a shorter wavelength in the far-IR, has a flatter SED from mid- to far-IR, and has a less prominent silicate absorption feature indicating less extreme obscuration towards the starburst region. 
Perhaps more interesting is the result that 8/33 of our sources have Cirrus-type SEDs. This suggests that cool, quiescently star-forming galaxies can make up a significant fraction ($\sim$25\%) of the submm population. Half of the cirrus-type sources (4/8) lie at moderate redshift (z$<$1) and so could be similar to the cold SN host galaxies reported in Clements et al. (2005). One, however, SHADES-SXDF 14, lies at z=2.24, suggesting that this kind of cool dust galaxy has a role even at high redshift, though there remain some issues with this identification (see Appendix A). 

Our approach of fitting SEDs to individual objects highlights the variety of submm SEDs in this population. Approaches that obtain a generic SMG SED by fitting a single SED-template to photometry from a range of objects eg. Pope et al. (2006) will mask this variation despite the better wavelength sampling coming from combining sources at a range of redshifts providing a better constraint on this generic SED.

\subsection{The Role of AGN}

Our SED fits show little evidence for a significant bolometric luminosity contribution to the SMGs from AGN in the radio-detected SXDF SHADES sources. One source (SHADES-SXDF 21) may have its optical-near-IR SED dominated by an AGN, though an Sbc-type template is as good a fit. One other source (603) is a composite in the far-IR, including Arp220 starburst and AGN dust torus components where the AGN is the dominant source of luminosity by a factor of $\sim$2. There is some evidence for a contribution to the mid-to-far-IR SED coming from an AGN in five other sources (2, 4, 30, 37 and 69, though 69's identification must be treated with caution (see Appendix A)), but the AGN  is not the dominant component.
X-ray (Alexander et al., 2005) and rest-frame optical spectroscopic observations (Swinbank et al. (2004); Takata et al. (2006), though this latter sample was biased towards AGN selection) found that a significant fraction of SMGs contain AGN (75\% from the X-ray data, 40 --- 65\% from rest-frame optical spectroscopy). Pope et al. (2007), however, find fewer clear AGN hosting SMGs using mid-IR spectroscopy. Our broad-band optical/Spitzer observations cannot look for the presence of an AGN, but can examine the role of any AGN in the energetics of the objects. Our results are thus consistent with Alexander et al.'s conclusion that despite the presence of an AGN in many SMGs, they are nevertheless almost entirely powered by starbursts, with AGNi contributing to the bolometric luminosity at the $\sim 10\%$ level. This conclusion is in line with Alexander et al. (2005)'s analysis of the GOODS-N SMGs. We find no evidence to suggest the presence of a large fraction of compton thick AGN in our SMGs. Such objects have been found elsewhere in the SWIRE survey (Polletta et al., 2006), but only one of the sources discussed here is a possible Quasar 2 candidate. Our data thus supports the view that AGN, compton thick or otherwise, play only a limited role in the bolometric luminosity of SMGs.

\subsection{The Stellar Mass of the Host Galaxies: Evidence for Downsizing?}

Examination of the stellar mass of SMGs allows some insight into their broader evolutionary role. Are they, for example, massive systems, and does the nature of an SMG change with redshift? Determinations of stellar mass are usually based on the rest-frame K-band luminosity, but the exact conversion from K to stellar mass is dependent on the details of the star formation history of each object (see eg. Borys et al., (2005), and Dye et al., (2007)). Given the paucity of data points for most of our sources, and the number of parameters already fitted to them, we here only consider the rest-frame K-band luminosity, derived from interpolation between our observed fluxes once redshift has been taken into account, as a surrogate for the stellar mass, and see how this varies with redshift. This is plotted in Figure 4, along with similarly derived K-band luminosities from the significantly deeper GOODS-N Spitzer data. 

The K-band luminosity of our SMGs shows a clear decline from z=3 to z=0, showing that far-IR-luminous SMGs are likely to have smaller stellar masses at lower redshifts. We find, for example, 9 objects with $L_K > 2 \times 10^{11} L_{\odot}$ between z=2 and z=3, but only one between z=0 and z=2. The comoving volume between z=0 and 2 is 75\% that between z=2 and 3, so we would expect 7$\pm$3 $L_K > 2 \times 10^{11} L_{\odot}$, on the basis of Poisson statistics, between z=0 and 2. Instead we see only one. This would seem to be evidence for 'downsizing' in the SMG population, ie. the massive starbursts that power the submm activity are taking place in lower mass systems at lower redshifts and, conversely, in higher mass systems at higher redshifts. The term 'downsizing' (Cowie et al., 1996) has generally been used to refer to the observation that more massive galaxies seem to form their stars at earlier epochs and on shorter  timescales (Thomas et al., 2005). This effect has been seen in both spiral galaxies (Nelan et al., 2005; Thomas et al., 2005) and in larger, more generic, galaxy samples from SDSS (Heavens et al., 2004; Jimenez et al., 2005). Since SMGs are often thought to be elliptical galaxies in the process of formation, our possible detection of downsizing is especially interesting. Indeed the overall downsizing behaviour of SMGs suggested here seems to match that of the massive galaxies studied in the Gemini Deep Deep Survey (GDDS, Juneau et al., 2005). Massive (M$_* > 10^{10.8}$) galaxies in GDDS were found to have a high, bursting, star formation rate at z$\sim 2$ which subsequently declines while intermediate mass systems ($10^{10.2}<M<10^{10.8}$) have peak star formation rates at z$\sim$1.5 which then decline more slowly. It should also be noted that, if this effect is real, then galaxy formation for the progenitors of SMGs must be sufficiently rapid to produce massive galaxies at redshifts $\sim$ 2.5.

Our putative detection of downsizing in the SMG population is complicated by the relatively bright flux limits of the SWIRE IRAC bands for our SHADES sources. However, the GOODS-N SMGs discussed by Pope et al. (2006), for which much deeper IRAC data is available, follow similar downsizing-like behaviour (see Fig. 4), while a separate and more sophisticated analysis of  the Lockman SHADES sources (Dye et al., 2007) also shows the effect.

\begin{figure}
\epsfig{file=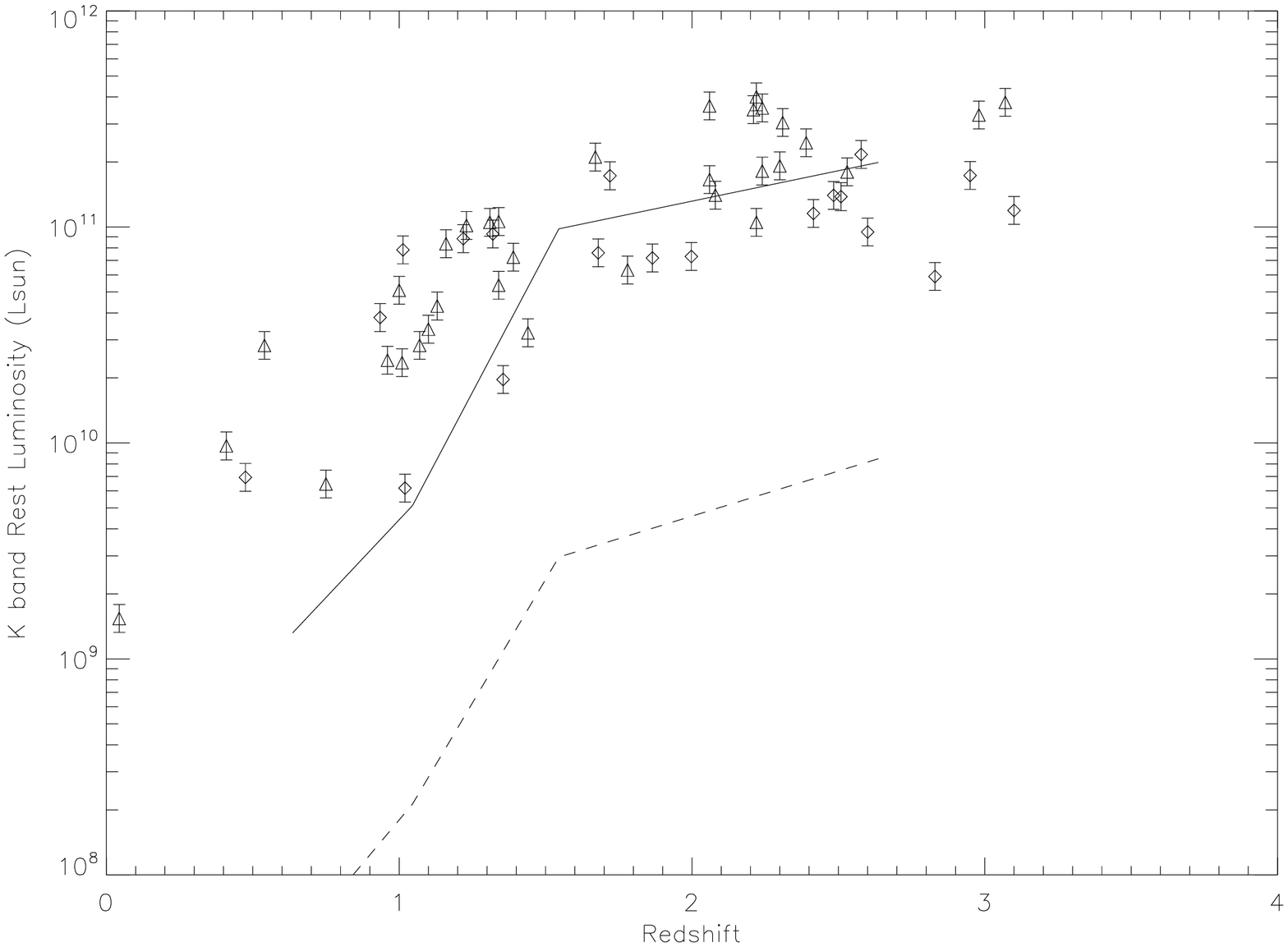,width=7cm, angle=90}
\caption{Rest-frame K band luminosity as a function of redshift for SMG galaxies 
illustrating apparent downsizing of SMG hosts. 
Triangles are sources from this paper, diamonds are sources from the GOODS-N 'supermap' (Pope et al., 2006). The lines indicate the detectable stellar mass as a function of redshift derived from the raw detection limits of the two surveys --- solid line for the SHADES/SWIRE sources, dashed line for the GOODS-N sources. Some of the SHADES/SWIRE sources lie beneath this line as a result of extrapolation from adjacent bands when a source is not detected in the band closest to the emitted-frame K band. Despite the GOODS-N survey having a much deeper flux limit than SWIRE, the downward trend of stellar mass with redshift is similar for GOODS-N and SHADES/SWIRE sources. This suggests the decline in K band luminosity, and thus stellar mass, with redshift is real rather than a result of the sources lying close to the flux limit,
suggesting that downsizing is taking place in the SMG population.}
\end{figure} 

\subsection{Physical Associations}

Our examination of ambiguously identified sources has revealed a number of cases where several objects appear to be physically associated with the SCUBA source. Whether this is as interacting pairs/groups or as close associations in some larger structure such as a cluster is uncertain. The clearest example of this is SHADES-SXDF 47, where three radio identified sources are candidate identifications, all lying at the same estimated redshift z = 1.3. SHADES-SXDF 27 has two non-radio IDed objects at z$_{phot}$=2.01 that appear to be associated with the far-IR source, while SHADES-SXDF 28 has two radio identifications both at z$_{phot}$=1.88. SHADES14 has two radio counterparts, but one of these has no optical or SWIRE detection. Given the redshift estimated for the other radio/SWIRE source in this system, z=2.24, it is quite possible that the companion to this source lies at the same redshift but is just too faint for us to see in SWIRE or SXDF. Pairs of radio sources or galaxies on a range of scales have been found for a number of SMGs (eg. Blain et al., 2004; Swinbank et al., 2006). The physical sepearations between these sources range in size from 8 to 135 kpc, and thus cover the distance scales expected for both interacting pairs of galaxies and galaxy clusters. It is thus seems likely that some SHADES sources are to be found in either interacting groups or in candidate high redshift clusters.

Is this true for the rest of the population? For the bulk of the SHADES sources, there are no clearly selected companions, on the basis of radio or 24$\mu$m emission. We thus investigate the possibility of associations by examining the photometric redshifts of all potential companions within a 10" radius of the sources. This radius is chosen on the basis of correlation analysis (Serjeant et al., 2007) which suggests the presence of correlations on these angular scales. We calculate the difference in estimated redshift, $\Delta z$ between each SHADES source and all its companions and compare this to twenty reference fields offset from the SHADES sources by up to $\sim$0.1 degrees. If there is a real correlation in redshift between the bulk of the SHADES sources and their nearby companions then there will be an enhancement of this statistic at zero redshift difference for the SHADES sources that is not apparent for the reference fields. Figure 4 shows this comparison. As can be seen we find no enhancement in the number of objects with $\Delta z = 0$ for the SHADES SMGs relative to the reference fields. A Kolmogorov-Smirnoff test comparing the $\Delta z$ distributions for SHADES sources and reference fields finds only a 40\% probability that they are drawn from different distributions. By adding fake companions to this comparison we find that a 95\% or higher significance difference between the $\Delta z$ distributions for SHADES sources and reference fields can be detected if $\sim$1/5 of our SHADES sources (ie. 7 objects) have close companions with redshift differences less than 0.15.

We thus conclude that there is no evidence that the bulk of SHADES sources have an unusual number of close companions at the same estimated redshift, based on potential companions detected by SWIRE. The small number of cases where close companions are detected, such as SHADES-SXDF 47, are thus rather unusual and may warrant further investigation. A thorough investigation of the clustering properties of SHADES SMGs is underway (van Kampen et al., in preparation).

For those sources where there appears to be a real physical association, such as SHADES-SXDF 47, there is a possibility that the submm flux might come from more than one object. However, we here assume that the submm flux comes solely from one source. This is a simplifying assumption, removing a complicating additional free parameter from our analysis, but there are a number of precedents to suggest that far-IR/submm flux in well separated merger or cluster companions is not evenly distributed between them. In the local universe examples include the ULIRG IRAS 09111-1007 which comprises two components separated by $\sim$40kpc. These separate components  have a 4 to 1 ratio of submm flux (Khan et al., 2005). In the more distant universe SMA observations have shown that at least one SMG with two physically associated radio IDs has submm emission from only one of these components (Younger et al., 2007). We thus conclude that ascribing the submm emission of physically associated sources to just one of the plausible IDs is the most reasonable approach in the absence of evidence to the contrary.

\begin{figure}
\epsfig{file=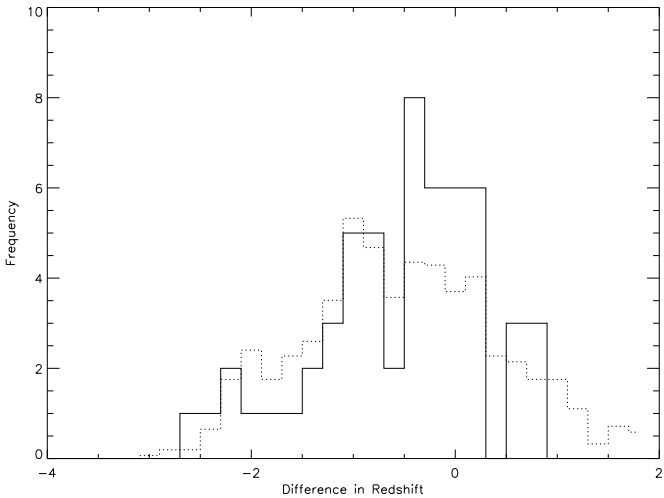,width=7cm, angle=90}
\caption{Searching for Physical Companions. 
Histogram of the redshift differences between SHADES sources and their companions within 10" (solid line) compared to redshift differences between SHADES sources and objects in 20 similar fields offset from the SHADES source by up to $\sim$0.1 degree (dotted line). As can be seen there is no significant difference between the two plots. We thus conclude there is no evidence for physical associations between generic SHADES sources and the general SWIRE galaxy population, even though there are clear examples of associations in specific cases.}
\end{figure} 







\section{Conclusions}
We have examined the counterparts to radio and/or 24$\mu$m selected identifications to SHADES SXDF SMGs in the SWIRE mid/far-IR and SXDF optical surveys. We find optical and/or SWIRE sources that correspond to 32 SHADES SMGs. Six of the radio-identified SHADES SMGs in the SXDF discussed in Ivison et al. (2006) are found to have neither optical nor SWIRE counterparts. We use the optical to mid-IR fluxes obtained for the 32 cross-identified SMGs to obtain both photometric redshift estimates and template fits to the optical-to-far-IR SEDs of the SMGs. We find that the SMG population revealed here is diverse in both redshift and in SED-type. The redshifts found range from $z=0.05$ to $z =3.27$, with far-IR luminosities ranging from 10$^{10.25}L_{\odot}$ to 10$^{13.26}L_{\odot}$. The majority of SMGs, though, are of ULIRG-type far-IR luminosity and lie at redshifts 1.5--2.5. Most ($\sim$60\%) of our SMGs have their SEDs fitted by an Arp220-type cool starburst ULIRG template, but there are significant numbers ($\sim$20\% each) that appear to be warmer M82-type starbursts and cold cirrus-type galaxies. The appearance of several cirrus-type sources in this survey supports Efstathiou \& Rowan-Robinson (2003)'s suggestion that some SMGs have a much colder dust temperature than local ULIRGs. This result is also supported by the detection that some z$\sim$0.5 SN1a host galaxies are submm bright but have cold dust temperatures (Clements et al., 2005; Farrah et al., 2004).

Our SED templates also allow us to assess the energetic contribution of any AGN that might lie in the SMG. We find no evidence that AGN make a significant contribution to the bolometric output of the SMGs in this sample, agreeing with the conclusions of Alexander et al. (2005) and Pope et al. (2006).

We find that lower redshift SMGs appear to have lower rest-frame K-band luminosities, suggesting that galaxy 'downsizing' (Thomas et al., 2005) is at work in the SMG population. This is an expected result in the generally assumed picture of SMGs being a stage in the evolution of elliptical galaxies.

Several of our SMGs are found to either be multiple sources with two or more components, or to lie in groups or clusters. However an analysis of the photometric-redshift distributions of all sources within 10" of SMGs finds no statistically significant number of companions compared to non-SMG objects.

This work represents the first stage of analysing the mid-to-far-IR SEDs of the SHADES submillimetre galaxies. The SHADES field in Lockman was observed by the Spitzer GTO teams with somewhat deeper integrations and is being discussed elsewhere (Dye et al., 2007) while specific subpopulations such as NIR-selected sources are also being investigated (Takagi et al., 2007). Examination of Spitzer counterparts to non-radio-detected sources in the SXDF and Lockman SHADES fields is also underway (Oliver et al., in preparation; Dye et al., in preparation), along with stacking analysis to produce population averages for undetected sources (Serjeant et al., in preparation).
\\~\\
{\bf Acknowledgements}
\\~\\
Thanks to Jeff Wagg and Marcos Trichas for useful comments. DLC is funded by PPARC/STFC, IRS is supported by the Royal Society. The JCMT is supported by the United 
KingdomÕs Science and Technology 
Facilities Council (STFC), the National Research Council Canada 
(NRC), and the Netherlands Organization for Scientific Research (NWO);  
it is overseen by the JCMT Board.  We acknowledge funding support from PPARC/STFC, NRC and NASA. The authors would like to thank the staff at the JCMT for their typically excellent support work. This research has made use of the NASA/IPAC Extragalactic Database (NED) which is operated by the Jet Propulsion Laboratory, California Institute of Technology, under contract with the National Aeronautics and Space Administration.

\begin{appendix}
\section{Notes on Specific Sources}

We here provide postage stamp images and describe the details of specific sources where greater explanation is needed concerning their photo-z, SEDs or identification. All P values quoted here come from Ivison et al. (2007).
\\~\\
{\bf SHADES-SXDF 5} This object appears in the optical as a complex chain of sources extending about 12" EW and about 5" NS. There is one radio source within this chain coincident with a SWIRE source. This SWIRE source is bright at 24$\mu$m with a P statistic of 0.008, indicating that it is very likely to be related to the SHADES source. This does not necessarily mean that the rest of the structure is not physically associated with the SHADES source. This object is also coincident with a 160$\mu$m source, and as such is one of only three 160$\mu$m detected objects in our sample.
\\~\\
{\bf SHADES-SXDF 6} This source is similar to SHADES 5, in that it appears in a complex group of optical and SWIRE sources. There are three radio sources close to the SHADES position. All radio sources have coincident 24$\mu$m sources. There is also a possible optical/SWIRE companion to the radio source nearest the nominal SHADES position.  We extract fluxes for all of the candidate identifications and list them in Table 1 as follows: 6 is the closest radio ID, with 602 and 603 being the radio IDs lying at increasing distances, while 601 is the closer non-radio ID. Observations with the SMA have been used to determine which of these sources is responsible for the submm emission (Dunlop et al., private communication). These show that 603 is the correct identification.
\\~\\
{\bf SHADES-SXDF 12} This source has no counterpart in the SWIRE catalogs, but it has a clear match between the radio ID and an optical source in the SXDF catalogs.
\\~\\
{\bf SHADES-SXDF 14} There are two radio sources associated with this SHADES source. The closer radio source (6" separation) is marginally brighter in the radio than the more distant source (9"), giving a better P$_{radio}$ of 0.04 versus 0.109 (Ivison et al., 2006) for the more distant source. However, the closer radio source lacks an optical or SWIRE counterpart. We here assume that the identification with a SWIRE source is correct for the rest of this paper, but this identification is not secure and should be treated with caution.
\\~\\
{\bf SHADES-SXDF 21} This source is the brightest optical and mid-IR source in our list of identifications. It is also strongly detected at 70 and 160 $\mu$m. The optical image shows a bright extended galaxy which is clearly also detected in the SWIRE bands, with the radio position lying at the centre of this source. The submm position, in contrast, lies somewhat to the east of the radio and SWIRE positions, but this is a fairly faint submm source for SHADES, and the shift in submm position relative to the radio/SWIRE positions is fully consistent with known pointing effects on faint sources (see eg. Eales et al. 2000). The optical identification for this source is the local UV excess galaxy KUG 0215-053 (Miyauchi-Isobe \& Maehara, 1998) discovered in the second KISO survey. Examples of such low redshift SMGs exist elsewhere (eg. CUDSS3.8, Clements et al. 2004). Chapman et al. (2005) have suggested that some such cases are the result of the low redshift object lensing a background high redshift SMG. We find that the SED of this object is  entirely consistent with a low redshift cirrus origin for the far-IR and submm emission. In this case at least, the source appears to be a genuine low z galaxy with prominent cold dust emission. A spectroscopic redshift is available for this source, z=0.044, which is consistent with the photo-z estimate of z=0.05. We have used the spectroscopic redshift for the SED fits.\\~\\
\\~\\
{\bf SHADES-SXDF 24}
There are two possible radio identifications for this source. The closest radio identification has a P$_{radio}$=0.014 (Ivison et al., 2006) but no optical or SWIRE counterpart. A more distant possible radio identification with P$_{radio}$= 0.047 has a matching SWIRE and optical counterpart including a 24$\mu$m detection with P$_{24}$ = 0.034. We extract optical and SWIRE fluxes for this counterpart and examine the resulting SED, which produces a reasonable fit to the submm emission. Nevertheless, the optical and SWIRE blank field radio ID may still be the correct identification.
\\~\\
{\bf SHADES-SXDF 27}
The radio ID for this source has two non-radio detected SWIRE companions. All three turn out to have the same photo-z estimate,  z$\sim$ 2.0, with matching far-IR SEDs. This would thus seem to be a high redshfit cluster or group. We cannot say which of them is the SCUBA source, or whether the SCUBA source contains a contribution from more than one of these objects. The SED plotted is based solely on the optical and SWIRE fluxes for the radio ID. 
\\~\\
{\bf SHADES-SXDF 28} There are two radio sources of similar brightness in radio, optical and SWIRE bands at similar separations (3.3" and 3.8") from the nominal SCUBA position, though the closer source is somewhat redder in its SED. The photo-z process finds similar properties for each object, both of which appear to lie at z$\sim$1.16. Both are potential identifications,
and there is also the possibility that these objects are physically associated. For the remainder of this paper we treat the nearer of the radio sources as the correct identification, but our results are not significantly different if the second source is the correct identification. The SED plotted is based solely on the optical and SWIRE fluxes of the nearer of the radio sources.
\\~\\
{\bf SHADES-SXDF 29}
This SCUBA source initially appears associated with one of two neighbouring objects. We have extracted fluxes for both of these (in Table 1, 29 indicates the radio associated source, 2902 the non-radio companion). Photo-z estimates for their redshifts are very similar, indicating z=0.175 Sab galaxies. However, the mid-to-far SED fit fails for both these objects, with problems at all wavelengths. This would seem to suggest that the SCUBA source is associated with neither of the two low redshift galaxies. Closer inspection of the SWIRE images suggests that there may be a third source in this complex, listed as 29+ in Table 1, with the 3.6$\mu$m image showing an extension to the south of the eastern-most of the two possible optical associations that is not apparent in the optical. This 3.6$\mu$m extension coincides with the radio and 24$\mu$m source positions. However, because of blending with the two bright optical sources it has proven impossible for us to extract reliable optical photometry for this possible third source. Of all the SHADES objects discussed here, this is the strongest candidate gravitational lensing system, as has been suggested by Chapman et al. (2002) for some low z SMGs. More observations, for example submm interferometry, will be needed to confirm this. The source parameters given in Table 3 for this source are our best attempt at a photometric redshift using only the SWIRE fluxes for the southern extension and should be treated with caution. 
\\~\\
{\bf SHADES-SXDF 31} The radio and SWIRE counterpart to this SHADES source lies outside the usual 6" error circle. Nevertheless the identification has a low probability of being a random association thanks to both the radio and 24$\mu$m emission (P$_{radio}$=0.039, and P$_{24}$ = 0.025). However, the SED fits to this source suggest a different interpretation. Despite a good photo-z fit to an Sab galaxy at z=0.62, the mid-to-far-IR SED fit is poor for this source, failing at both 24 and 3.6 $\mu$m. We thus attempt to match the SCUBA source to the nearer SWIRE source without radio counterpart designated 3102 in table 1. This does much better, yielding a z=2.21 Sab galaxy and a good mid-to-far-IR fit. We thus conclude that the radio and 24$\mu$m emission from our original identification is not associated with the submm emission detected by SCUBA, and that 3102 is likely to be the correct identification.
\\~\\
{\bf SHADES-SXDF 35} This source has a spectroscopic redshift of z=1.255 (Dunlop et al., private communication) which matches well with our photometric estimate of z=1.23.
\\~\\
{\bf SHADES-SXDF 47} There are three radio sources associated with this SHADES source, all of which have SWIRE counterparts. The closest radio counterpart is the brightest at SWIRE wavelengths and has the best SWIRE P$_{24}$ value (0.015 vs. 0.048). However, the brightest, but more distant, radio source has the best radio P value (0.015 vs. 0.018) and we treat this as the primary identification (4701 in Table 1, with the other sources listed as 4702 and 4703 (the 24$\mu$m source)). We extract optical and SWIRE fluxes for all these sources and conduct photo-z and SED fits. All the sources have the same photo-z estimated redshift, z $\sim$1.3. They would thus appear to be physically associated. The SED fit to the optical and SWIRE fluxes of the original association (given under 4701 in table 1), along with the submm flux for this source, is good, so there is no reason to suspect that the submm emission is associated with either of the other sources. 
\\~\\
{\bf SHADES-SXDF 52} There are two radio sources with similar radio fluxes associated with this object. The closest of them is the brightest in the optical and SWIRE bands, and is also detected at 24$\mu$m. We thus select this source as the initial identification, with the second source designated 5202 in Table 1. However, the mid-to-far-IR SED fit for the primary identification, with a photo-z of z=0.36, has serious problems with the mid-IR IRAC fluxes. The more distant radio ID, 5202 in Table 1, has a much better SED fit, indicating a z=2.98 (an alias favoured by the mid-to-far-IR SED over a better photo-z in the optical-near-IR of 1.97) Arp220-type mid-to-far-IR SED on top of a proto-elliptical optical SED. This second source would thus appear to be a better identification than the radio source closer to the nominal SCUBA position.
\\~\\
{\bf SHADES-SXDF 69} The radio source is offset from the nominal SCUBA position by 13" for this object, sufficiently far that the P calculation becomes unreliable. Nevertheless, the radio source has a bright 24$\mu$m counterpart, and is thus  potentially associated with the SCUBA source. Its P value at 24$\mu$m is 0.068, though, so the identification should be treated with caution.
\\~\\
{\bf SHADES-SXDF 71}
This source does not have a radio ID but is plausibly matched (P = 0.019) by Ivison et al. to a 24$\mu$m source.
\\~\\
{\bf SHADES-SXDF 74}
There is a separation of 3" between the identified optical/SWIRE source and the radio position for this object. This may indicate that the association between the two is weak, so this identification should be treated with caution. Nevertheless, the photo-z and mid-to-far-IR SED show no problems for this identification.
\\~\\
{\bf SHADES-SXDF 76}
The separation between the SCUBA position and the radio counterpart for this source is 8.5". Nevertheless the radio P value is 0.049 so it remains a likely association.
\\~\\
{\bf SHADES-SXDF 77}
There are two radio associations for this source. The more distant association (9.5") has a coincident 24$\mu$m source, with P$_{radio}$=0.093 and P$_{24}$=0.021. The closer (4") radio source, and its SWIRE counterpart has a P$_{radio}$=0.047 and a weak 24$\mu$m association with P$_{24}$ = 0.042. The closer association is labeled 77, the more distant 7702. The photo-z and SED analyses of the primary identification are reasonable, with a photo-z of z=1.21. Nevertheless we perform the same analysis on the alternative, more distant radio/24$\mu$m identification. This source turns out to have a photo-z of z=0.96 and to have a better match for its mid-to-far-IR SED, so we thus marginally favour the z=0.96 identification. 
\\~\\
{\bf SHADES-SXDF 88}
The separation between the SCUBA position and the radio counterpart for this source is 8.85". The radio P value is 0.091 but this is also associated with a 24 $\mu$m source with a P$_{24}$ of 0.079. The radio and 24$\mu$m combination suggests that this is the correct identification, but the P values indicate that this association should be treated with caution.
\\~\\
{\bf SHADES-SXDF 96}
There are two radio associations with this source. The more distant association (12.7") has a coincident 24$\mu$m source but we follow Ivison et al. (2007) in ascribing the identification to the closer (5.4") radio source since it has a lower P value (P$_{radio}$=0.039) than the more distant 24$\mu$m/radio identification (P$_{24}$=0.097). No problems emerge for this identification from the photo-z or mid-to-far-IR SED fitting.
\\~\\
{\bf SHADES-SXDF 119}
There are two radio associations with this source. The more distant association (8.4") has a coincident bright optical and SWIRE source with strong 24$\mu$m emission. This has similar radio P values to the nearer (6") radio source (0.056 vs. 0.043) but also a good 24$\mu$m P value (0.019). We thus follow Ivison et al. (2007) in ascribing the identification to the more distant 24$\mu$m source rather than to the   more nearby (6") radio source which has no corresponding Spitzer or optical emission. The photo-z and SED fitting for this identification are reasonable, supporting the idea that it is the correct identification.
\\~\\
Postage Stamp Images at optical $i$ band, 3.6$\mu$m and 24$\mu$m will be available in the published version of this paper and can be obtained by contacting the first author, D.L. Clements.

\end{appendix}

\end{document}